\documentclass{article}

\usepackage{epsfig}
\usepackage{amsmath}

\begin{document}

%

\setcounter{figure}{0} \setcounter{table}{0} \setcounter{footnote}{0} \setcounter{equation}{0}

\begin{center}
\noindent \\[0pt]
{\bf\Large (SC)RMI: \\
A (S)emi-(C)lassical (R)elativistic (M)otion
(I)integrator, to model the orbits of space probes around the Earth and other planets}\vspace*{1cm}
\end{center}

\begin{center}
\noindent \\[0pt]
{\Large S. Pireaux}\\[0pt]
Sophie.Pireaux@obs-azur.fr\\
D\'{e}partement ARTEMIS, Observatoire de la C\^{o}te d'Azur, GRASSE, FRANCE\\[0pt] 
\vspace*{0.25cm}

{\Large J-P. Barriot}\\[0pt]
Jean-Pierre.Barriot@cnes.fr\\
UMR5562 DTP, Observatoire Midi-Pyr\'{e}n\'{e}es, TOULOUSE, FRANCE\\[0pt]
\vspace*{0.25cm}

{\Large P. Rosenblatt}\\[0pt]
Pascal.Rosenblatt@oma.be\\
Observatoire Royal de Belgique, BRUXELLES, BELGIQUE\\[0pt]
\end{center}

\vspace*{2cm}

\begin{abstract}

Today, the motion of spacecrafts is still described according to the
classical Newtonian equations plus the so-called « relativistic corrections», 
computed with the required precision using the Post-(Post-) Newtonian formalism. 
The current approach, with the increase of tracking precision 
(Ka-Band Doppler, interplanetary lasers) and clock stabilities (atomic 
fountains) is reaching its limits in terms of complexity, and is furthermore 
error prone. In the appropriate framework of General
Relativity, we study a method to numerically integrate the native
relativistic equations of motion for a weak gravitational field, also taking into
account small non-gravitational forces. 
The latter are treated as perturbations, in the sense that we assume that both the local 
structure of space-time is not modified by these forces, and that the unperturbed satellite 
motion follows the geodesics of the local space-time. 
The use of a symplectic integrator to compute the unperturbed geodesic motion
insures the constancy of the norm of the proper velocity quadrivector. 
We further show how this general relativistic framework relates to the classical one.

\vspace*{0.5cm}
\end{abstract}


\section{The classical approach: \textit{GINS}}

\begin{figure}[t]
\includegraphics[width=.7\textwidth]{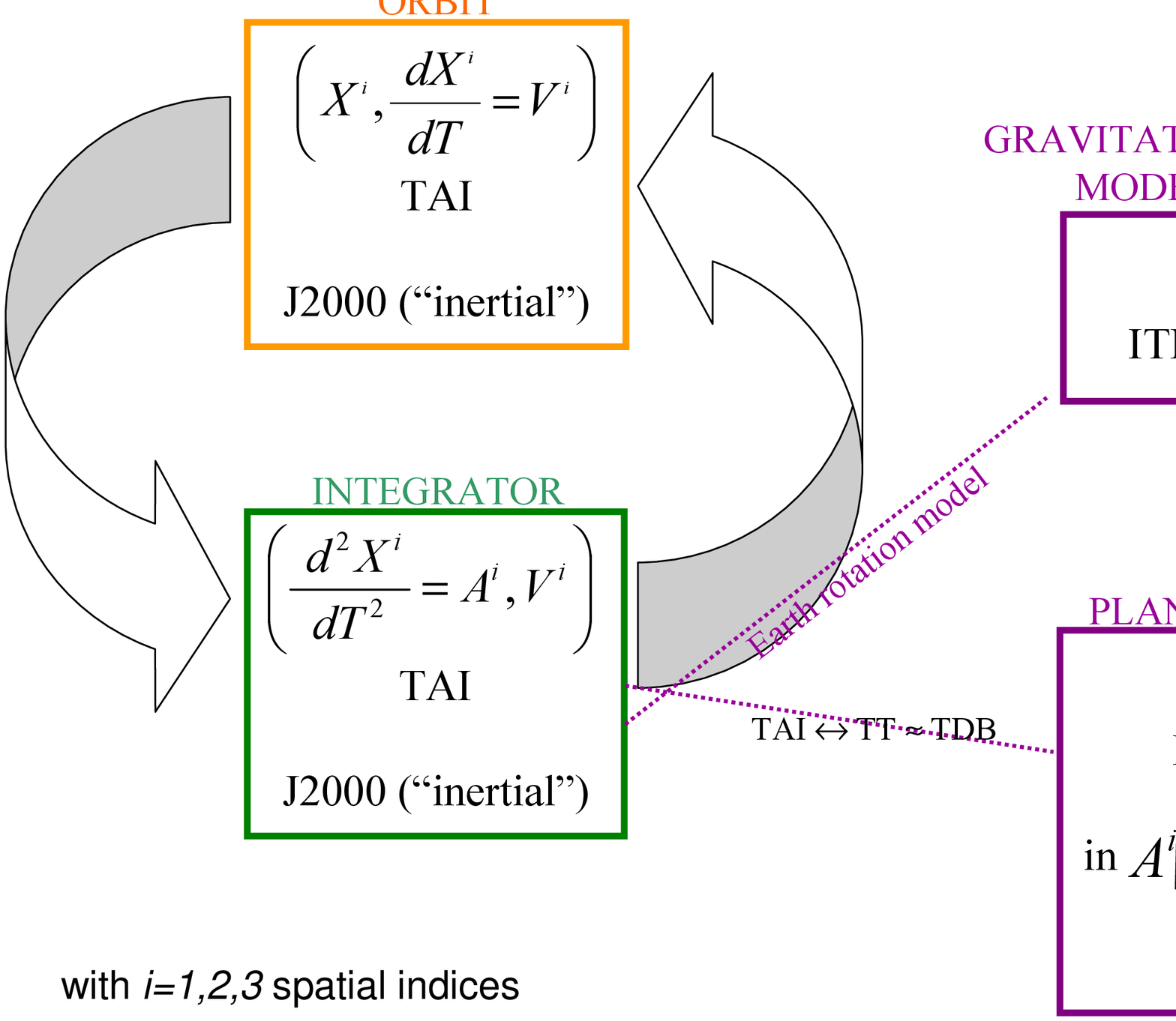}
\caption[]{GINS. \newline
		This diagram summarizes the GINS orbitography software approach.
		The orbit of the satellite is given in terms of 3-dimensional position and velocity
		vectors, $X^i$ and $V^i=dX^i/dT$ respectively, in the J2000 quasi-inertial Earth frame, 
		with respect to International Atomic Time (TAI). 
		It is the 3-dimensional acceleration, $A^i=d^2X^i/dT^2$, and velocity vectors
		which are integrated with respect to this time. The central gravitational potential can be either the Earth (E)
		potential given for example by the GRIM5-S1 model provided in the International Terrestrial Reference 
		System (ITRS) or any central body planetary potential, completed by the potential of other celestial bodies. 
		Velocities and positions of planets (P), that is $x^i_P$ and $v^i_P$, provided in terms of TDB 
		(``Temps Dynamique Barycentrique'') in the barycentric reference frame by planetary ephemeris (such as DE403) 
		are needed to 
		compute the Newtonian contribution from planets as well as relativistic accelerations (Geodesic Precession (GP)). 
 		Additional space (Earth or central body rotation model) and time (TAI $\leftrightarrow$ TDB, with TDB assumed 
 		in GINS to be equal to the Terrestrial Time (TT)) are needed to transform between the different reference frames. }
\label{GINS}
\end{figure}

In today's planetary orbitography softwares, as in GINS (G\'{e}od\'{e}sie
par Int\'{e}grations Num\'{e}riques Simultan\'{e}es, developed by 
CNES\footnote{Centre National d'Etudes Spatiales, France} and 
GRGS\footnote{Groupe de Recherche en G\'{e}od\'{e}sie Spatiale, France}), 
the motion of spacecrafts is still described according to the classical Newtonian
equations plus the so-called ``relativistic corrections'', computed with the
required precision using the Post-(Post-) Newtonian formalism.
\newline Hence, it is the 3-dimensional acceleration vector ($i=1,2,3$), 
which is numerically integrated with respect to coordinate time $T$,
\begin{equation}
\frac{d^{2}X^{i}}{dT^{2}}=\left. 
\begin{array}[t]{l}
-\frac{\partial W}{\partial X^{i}}\smallskip \\ 
- K^{i}\smallskip \\ 
+
\text{\textit{general relativistic corrections.}}
\end{array}
\right.  
\label{newton_equation_of_motion}
\end{equation}
The gravitational potential $W$ includes not only the central planetary
potential model but also the Earth-tide potential (due to the Sun and Moon, corrected
for Love number according to the frequencies, ellipticity and polar tides),
ocean-tide potential and Newtonian-perturbation potentials from other solar
system bodies. The atmospheric drag, the radiation pressure (solar
radiation, Earth albedo, thermal emission) are the non-gravitational
perturbations considered, $K^{i}$. The orbitography software GINS also includes, as
relativisitic corrections, the Schwarzschild, geodesic and Lense-Thirring
precessions \cite{descriptif GINS}.
Figure \ref{GINS} summarizes the GINS
approach. It is a generic one which can be applied to orbitography
around any central planet.

When analysing satellite data for geophysical reduction, General Relativity
not only plays a role in computing the precise satellite orbit through equation (\ref
{newton_equation_of_motion}). Indeed, general relativistic corrections are
also applied on measurements, because electromagnetic signals travel in
curved space-time, and are thus deflected and delayed. Also, the planetary
potential model is described in the planetary crust frame, rotating with the
central planet; while the satellite motion is described in the planetary
quasi-inertial frame, non rotating with the planet. Both frames are linked
through the planetary rotation model. 
This model must include the
Earth relativistic geodesic precession, 
\cite{IAU2000 resolutions}, \cite{explanatory on IAU2000 resolutions}, i.e. $\sim $2 arcseconds/century
mostly due to the Sun static gravitational field, a general relativistic
correction steming from the fact that the Earth frame is not an inertial frame but
moves in the gravitational field generated by other solar system bodies.

\noindent Furthermore, any reference frame transformation (geocentric to barycentric
and vice-versa, needed to get positions and velocities of solar system
bodies from ephemerides) is a 4-dimensional space-time
transformation in General Relativity. This means adding more relativistic
corrections to the corresponding Newtonian 3-space transformations and
abolishing the status of an absolute time. Time is intricately related with
space, and time is relative to a reference system.

\section{Motivations}

\subsection{Relevance of relativistic effects}

\begin{center}
\begin{table*}[f]
\begin{tabular}{lll} \hline
\textbf{Cause (m/s}$^{2}$\textbf{)} & \textbf{LAGEOS 1} & \textbf{CHAMP} \\
\hline 
\textbf{G:} Earth monopole & 2.8 & 8.6 \\ 
\textbf{G:} Low order geopotential harmonics (eg. l=2,m=2) & 6.0 10$^{-6}$ & 6.4 10$^{-5}$ \\ 
\textbf{G:} High order geopotential harmonics (eg. l=18,m=18) & 6.9 10$^{-12}$ & 9.4 10$^{-7}$ \\ 
\textbf{G:} Moon & 2.1 10$^{-6}$ & 7.9 10$^{-7}$ \\ 
\textbf{G:} Sun & 9.6 10$^{-7}$ & 2.7 10$^{-7}$ \\ 
\textbf{G:} Other planets (eg. Venus) & 1.3 10$^{-10}$ & 9.8 10$^{-13}$ \\ 
\textbf{G:} Indirect oblation (Moon-Earth) & 1.4 10$^{-11}$ & 1.4 10$^{-11}$ \\ 
\textbf{G:} General relativistic corrections (total) & 9.5 10$^{-10}$ & 1.7 10$^{-8}$ \\ 
\textbf{NG:} Atmospheric drag & 3 10$^{-12}$ & 3.5 10$^{-7}$ \\ 	
\textbf{NG:} Solar radiation pressure & 3.2 10$^{-9}$ & 3.2 10$^{-8}$ \\ 
\textbf{NG:} Earth albedo pressure & 3.4 10$^{-10}$ & 3.3 10$^{-9}$ \\ 
\textbf{NG:} Thermal emission & 1.9 10$^{-12}$ & 8.3 10$^{-9}$ \\
\hline
\label{Table_low_high_satellite}
\end{tabular}
%
      \parbox{14.5cm}{
	\caption{
	\small Gravitational and non-gravitational satellite accelerations.\newline
	Orders of magnitude of the gravitational (G) or non-gravitational (NG) accelerations
	in m/s$^{2}$ experienced by a high- or a low-orbit satellite (LAGEOS 1 or CHAMP respectively).
	Discrepancies between accelerations on LAGEOS 1 or on CHAMP due to planets (especially Venus) come 
	from the fact that the value was estimated at a different Julian Days, when the position of the 
	corresponding planet in the solar system was different. 
	The Laser GEOdynamics Satellite 1 was designed to calculate
	station positions (with a precision of about 1-3 cm), monitor tectonic-plate
	motion, measure the Earth gravitational field and Earth rotation. It is a
	passive (no onboard sensors/electronic, no attitude control) spherical
	satellite with laser reflectors, on a 5858 x 5958 km, i = 52.6${{}^\circ}$,
	orbit around Earth. The mission was launched in 1976 by the USA for an
	minimum lifetime of 50 years. The CHAllenging Minisatellite Payload is a
	German mission for precise gravity and magnetic field modeling
	including space and time variations. It holds laser reflectors, a GPS
	receiver, a drift meter, a magnetometer, a star sensor and accelerometers.
	It was launched in 2000 for a five-year mission on a near polar Earth orbit
	at an initial altitude of 454 km.
}
} 
      \label{Table_low_high_satellite}
\end{table*}
\end{center}
To illustrate the relevance of relativistic corrections on a
classical orbit, let us consider LAGEOS as an example of a high-orbit
satellite, at an altitude of about 6000 km, designed for geodynamic studies. As an
example of a low satellite, we take CHAMP, dedicated to the precise determination
of the Earth gravitational field, flying at a mean altitude of 450 km.
Table \ref{Table_low_high_satellite} illustrates our point.
Relativistic corrections are crucial for a high orbit, because they are
of the same order of magnitude as the most important non-gravitational
effect, namely radiation pressure. As a low-orbit satellite is concerned, general
relativistic effects are comparable in order of magnitude to thermal
emission or radiation pressure. But, of course, at low altitudes, the
atmospheric drag is much more important.
\vspace{0.5cm} 

\begin{figure}[t]
\begin{center}
\includegraphics[width=.7\textwidth]{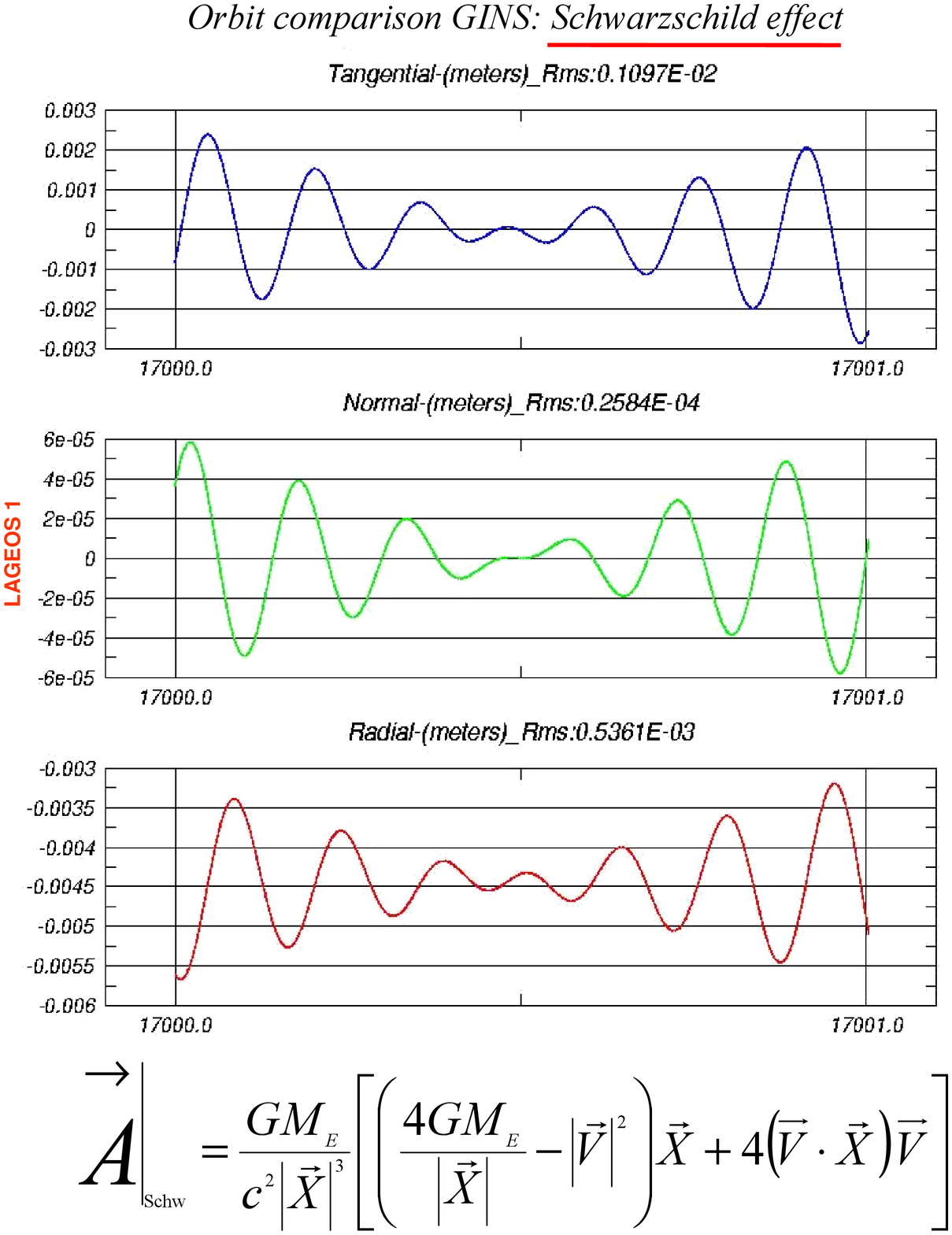}
\end{center}
\caption[]{Schwarzschild precession.\newline
The Schwarzschild general relativistic precession,
		$
%
%
%
		\left. \vec{A} \right| _{\text{Schw}}
		\equiv
		\left. d^{2}\vec{X}/dT^{2} \right| _{\text{Schw}}
		$,
		due to the Earth mass monopole, $GM_{E}$,
		is function of the satellite geocentric velocity, $\vec{V}$, and position, $\vec{X}$
		with respect to the Geocentric Coordinate Reference System (GCRS). 
		This effect is illustrated here with a one day data arc of satellite LAGEOS 1 by comparing two
		orbits generated by GINS, with or without that effect included. Tangential,
		normal and radial effects (rms) are provided in m.}
\label{Schwarz}
\end{figure}

\begin{figure}[t]
\begin{center}
\includegraphics[width=.7\textwidth]{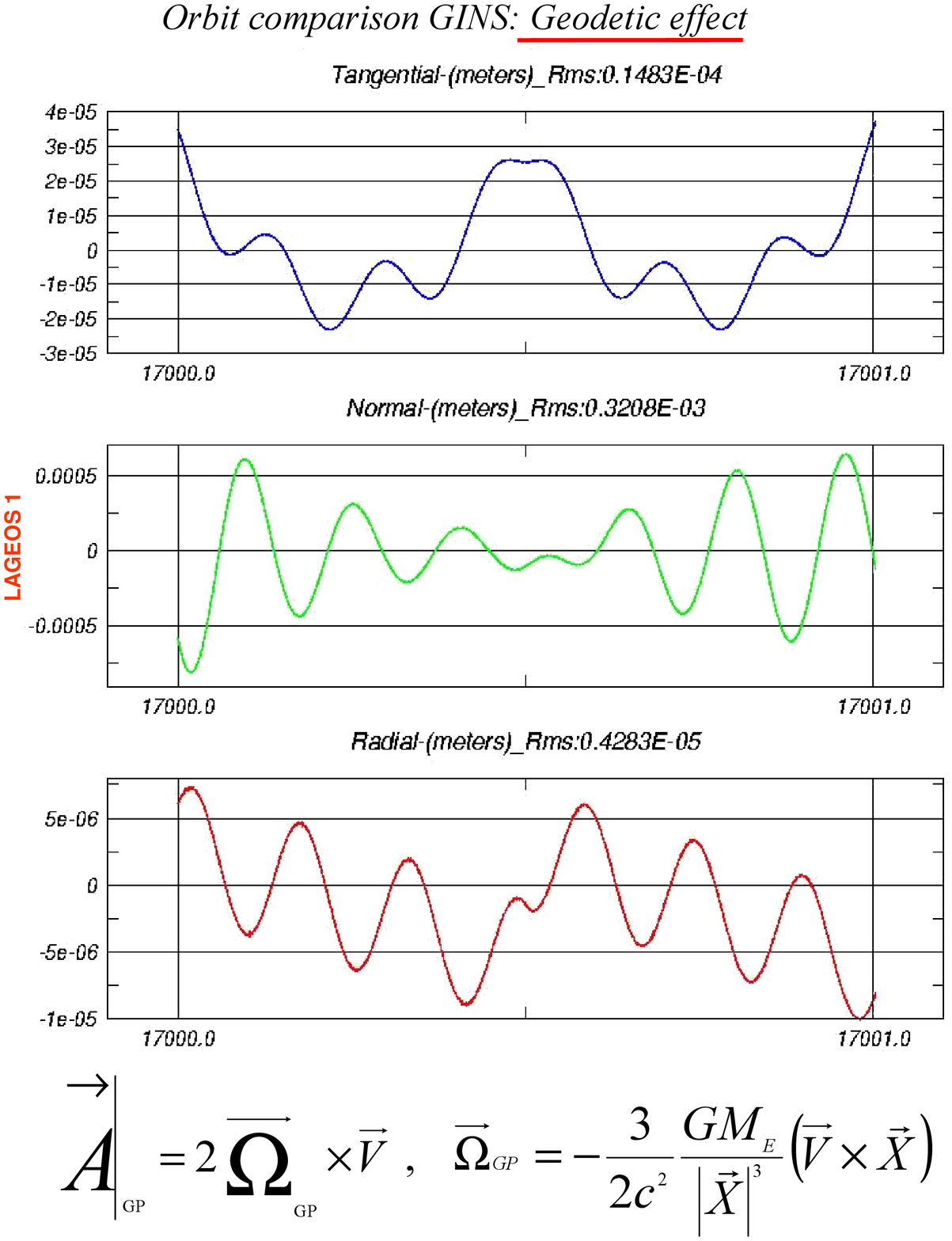}
\end{center}
\caption[]{Geodesic precession. \newline 
		The geodesic general relativistic precession, \newline 
		$
%
%
%
		\left. \vec{A} \right| _{\text{GP}}
		\equiv
		\left. d^{2}\vec{X}/dT^{2} \right| _{\text{GP}}
		$,
		due to the Earth mass monopole, $GM_{E}$,
		is function of the satellite geocentric velocity, $\vec{V}$, and position, $\vec{X}$
		with respect to the Geocentric Coordinate Reference System (GCRS). 
		This effect is illustrated here with a one day data arc of satellite LAGEOS 1 by comparing two
		orbits generated by GINS, with or without that effect included. Tangential,
		normal and radial effects (rms) are provided in m.}
\label{geodesic}
\end{figure}

\begin{figure}[t]
\begin{center}
\includegraphics[width=.7\textwidth]{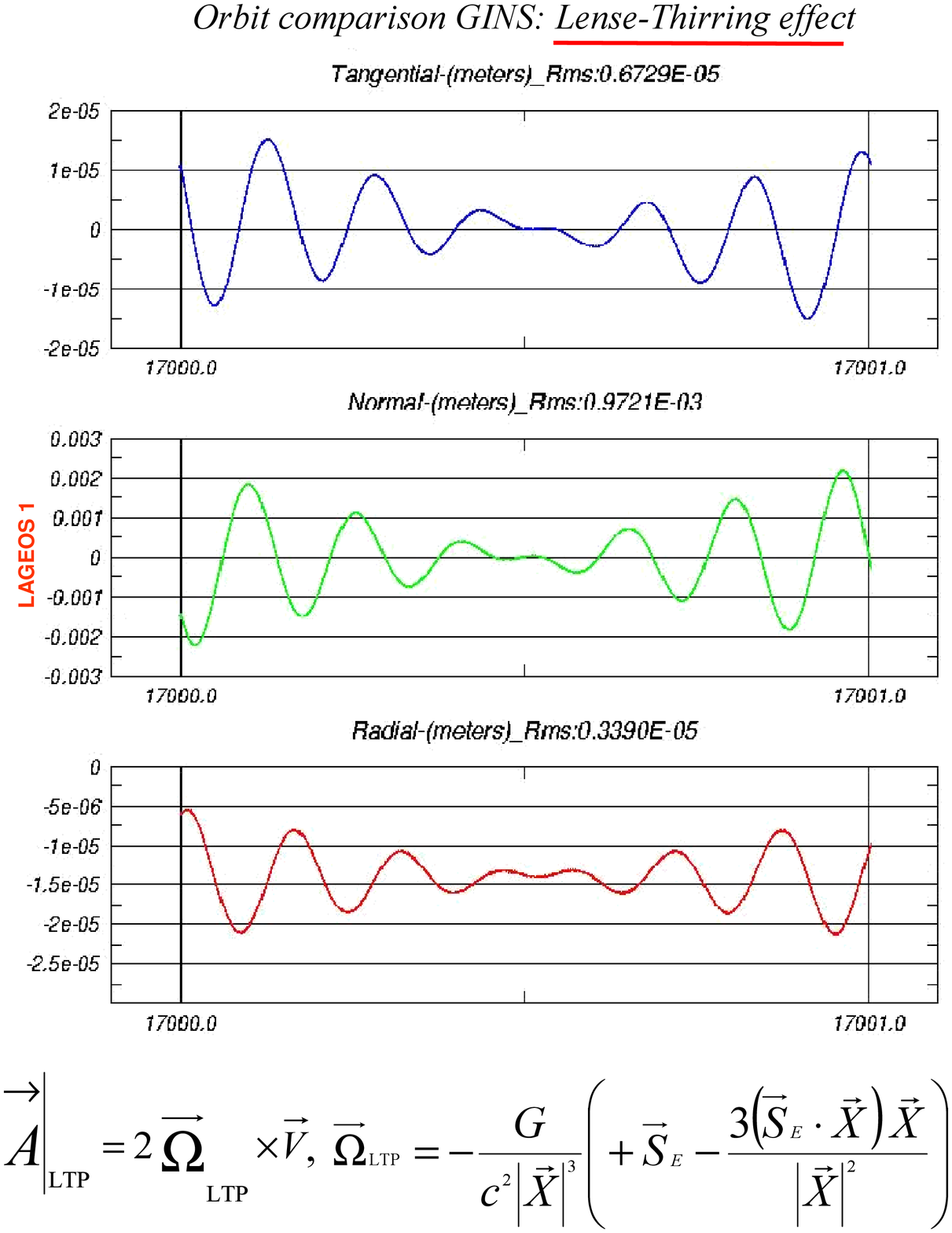}
\end{center}
\caption[]{Lense-Thirring precession. \newline
		The Lense-Thirring general relativistic precession, \newline 
		$
%
%
%
		\left. \vec{A} \right| _{\text{LTP}}
		\equiv
		\left. d^{2}\vec{X}/dT^{2} \right| _{\text{LTP}}
		$,
		due to the Earth spin dipole, $\vec{S}_{E}$,
		is function of the satellite geocentric velocity, $\vec{V}$, and position, $\vec{X}$
		with respect to the Geocentric Coordinate Reference System (GCRS).  
		This effect	is illustrated here with a one day data arc of satellite LAGEOS 1 by
		comparing two orbits generated by GINS, with or without that effect
		included. Tangential, normal and radial effects (rms) are provided in m.}
\label{Lense-Thirring}
\end{figure}

Figures \ref{Schwarz}, \ref{geodesic} and \ref{Lense-Thirring} detail the
relativistic corrections already included in the GINS software, and provide
orders of magnitude of the corresponding accelerations induced.\newline
\indent The Schwarschild precession is the most important one. It is associated with
the Earth mass monopole and mass multipoles, which deforms the space-time and hence affects the
satellite motion. It leads to a perigee advance of the orbit of the satellite
around the Earth, about 3 arcsec/year for LAGEOS 1.\newline
\indent Like the Schwarzschild precession, the geodesic precession is associated 
with the Earth mass. It is due to the fact
that the Earth ``warps'' the space-time around itself. Gravity Probe B, a NASA
mission, should be able to measure this effect, which is about 6.614 marcsec/year for gravity 
Probe B (on a polar orbit at 640 km), with a relative precision of 2 10$^{-5}$.\newline
\indent The Lense-Thirring general relativistic effect is associated with the Earth
spin dipole. As the Earth rotates, it ``drags'' the space-time, 
like a spining ball in molasses. The Lense-Thirring effect leads to an
additional perigee precession of the orbit of the satellite, plus a secular
drift of the 
\noindent ascending node. For LAGEOS 1, it is about 3 arcsec/century. For Gravity 
Probe B on a polar orbit, the Lense-Thirring and geodesic precessions are orthogonal.
This order of magnitude is comparable to the Schwarzschild effect due to the
degree 12 mass multipole moment of the Earth. Gravity Probe B, for which the
Lense-Thirring precession amounts to about 40.9 marcsec/year, should be able
to measure this general relativistic precession with a relative precision of 3 10$^{-3}$.\newline

\subsection{Precise geodesy for precise geophysics}

The present study is motivated by the fact 
\noindent that precise geophysics requires
precise geodesy.\newline
From the above paragraphs, we see that errors in relativistic corrections or
relativistic space-time transformations between reference frames lead to
a mis-modelling in the planetary potential and rotation model deduced from
precise orbitography. There is a real risk of polluting very weak geophysical
signatures, like the polar motion of Mars ($\sim 1$ m in amplitude at the
planet surface), or the nutations of its conjectured liquid core 
($\sim $ a few cm over an amplitude of $\sim $10 m), by unwanted
relativistic effects that are at the same period (typically one planetary
year, or 687 days 
\noindent for Mars), and, worse, that can be cumulative (ranging error 
up to or larger than 10 m coming from relativity over one spacecraft orbit around Mars, $\sim $150 minutes). 
Moreover, with the classical method, one correction can sometimes be counted
twice (for example, the reference frequency provided by GPS satellites
is already corrected for the main relativistic effect), if not forgotten.%

\subsection{Disadvantages of the classical method}

With the increase of tracking precision (32 GHz Ka/Ka-Band Doppler
radio tracking at the level of 1 $\mu$m/s with respect to a relative motion
Earth/spacecraft of 10 km/s, i.e. with a relative accuracy of $10^{-10}$),
active interplanetary laser tracking (at the level of 10 cm with respect to
a distance of $10^{8}$ km, i.e. with a relative accuracy of $10^{-12}$) and
clock stabilities (Allan deviation of $\sim $4 $10^{-14}\tau ^{-1/2}$ for
atomic fountains), the classical method is today reaching its limits in
terms of complexity. A (complete) review of all the corrections is needed
in case of any change in conventions (metric adopted), or if further precision
is gained in measurements.\newline
This is why, we suggest to use general relativistic mechanics from the begining, 
instead of Newtonian mechanics plus an increasing number of corrections. 

\section{The (Semi-Classical) relativistic approach: \\ \textit{(SC)RMI}}

\subsection{Relativistic equations}

In general relativistic mechanics, the relativistic equation of motion, when
non-gravitational accelerations (encoded in a 4-vector \cite{Lichnerowicz 1995 RG and em} 
$K_{\beta }$) 
are present, is
\begin{equation}
\frac{dU^{\alpha }}{d\tau }=-\Gamma _{\beta \gamma }^{\alpha }U^{\beta
}U^{\gamma }+K_{\beta }\left( G^{\alpha \beta }-\frac{U^{\alpha }}{c}%
\frac{U^{\beta }}{c}\right)   \label{generalized_equation_of_motion}
\end{equation}
\begin{equation}
\text{with }U^{\alpha }\equiv \frac{dX^{\alpha }}{d\tau }\text{,\quad }%
U^{\alpha }U_{\alpha }=c^{2}  \label{norm_conservation_equation}
\end{equation}
where $X^{\alpha =0,1,2,3}\equiv \left( c\cdot T,\ X^{i}\right) $ are the
space-time coordinates; $\Gamma _{\beta \gamma }^{\alpha }$ are the
Christoffel symbols; $G^{\alpha \beta }$, space-time metric; $c$ is the speed 
of light; and $\tau $ is the proper time. When $K_{\beta }=0$, equation (\ref
{generalized_equation_of_motion}) reduces to the geodesic equation of the
local space-time.

\subsection{(SC)RMI method}

We propose to numerically integrate the native relativistic equations
of motion for a weak gravitational field in the appropriate framework of
General Relativity, taking into account 
\begin{figure}[t]
\includegraphics[width=1.1\textwidth]{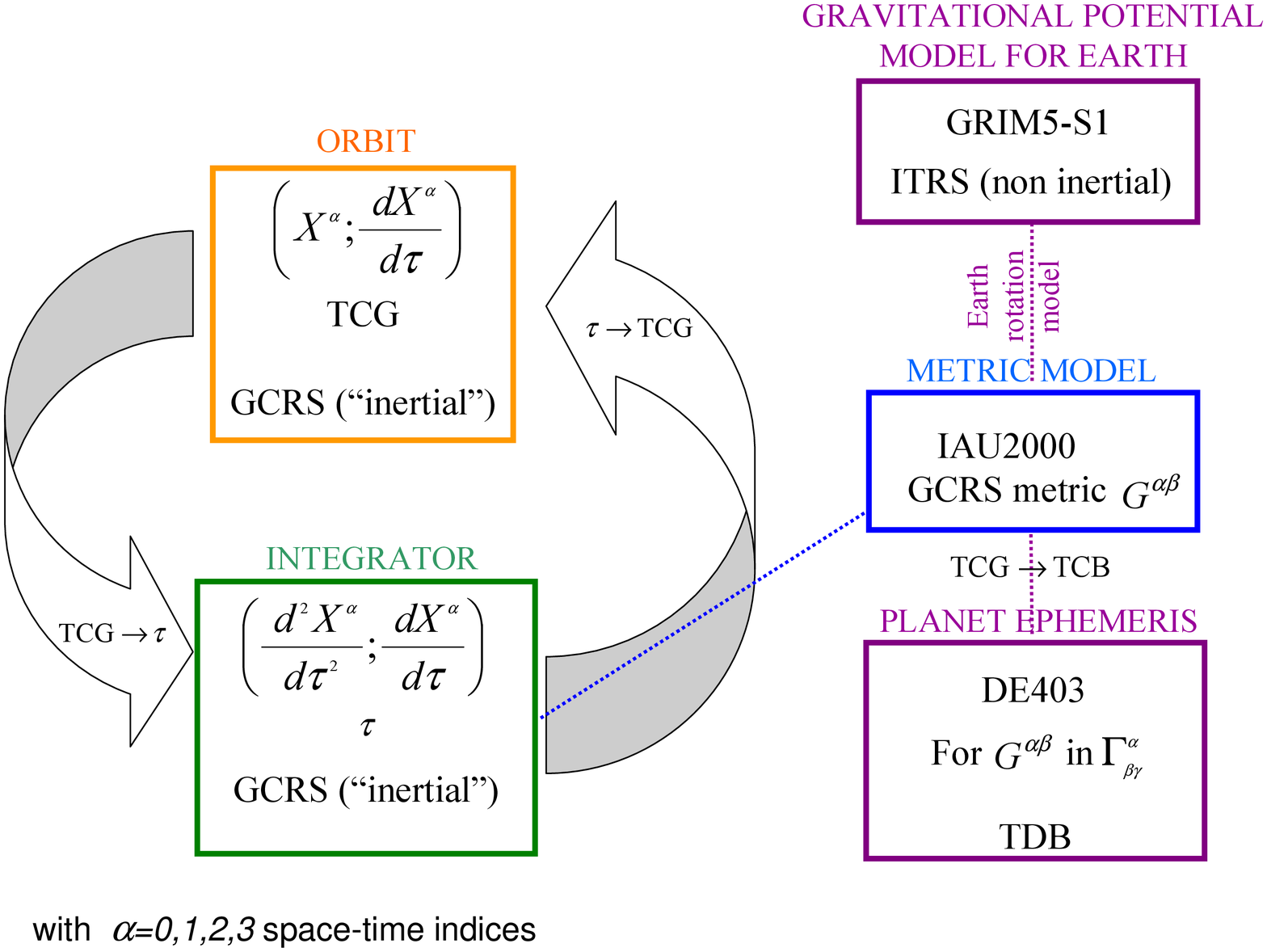}
\caption[]{(SC)RMI. \newline
		This diagram summarizes the (SC)RMI approach. The
		orbit is given in terms of 4--dimensional position and velocity 
		proper-vectors of the satellite, $X^\alpha$ and $dX^\alpha/d\tau$ respectively 
		(proper meaning with respect to the satellite proper time). 
		These quantities are referred to the J2000 Geocentric Coordinate Reference
		System (GCRS) Earth frame and its corresponding TCG time (``Temps de Coordonn\'{e}e G\'{e}ocentrique''). 
		It is the 4--dimensional proper-acceleration, $d^2X^\alpha/d\tau^2$, given by
		the relativistic equations of motion and the 4--dimensional proper-velocity
		vectors which are integrated with respect to proper time $\tau$,
		instead of coordinate time TCG. 
		To compute the relativistic equations, one needs to compute the GCRS metric, $G_{\alpha\beta}$, 
		and corresponding Christoffel symbols, $\Gamma^\alpha_{\beta\gamma}$, at a given space-time point, 
		according to the IAU2000 conventions. This requires a central planetary potential model 
		(such as GRIM5-S1 given in the International Terrestrial Reference System (ITRS) for the Earth), 
		plus planetary positions and velocities provided in ephemeris, such as DE403, in terms of TDB 
		(``Temps Dynamique Barycentrique'') or TCB (``Temps de Coordonn\'{e}e Barycentrique'' 
		related to the Barycentric Coordinate Reference System (BCRS)). 
		Additional relativistic (hence four-dimensional) space-time transformations are needed to transform 
		between the different relativistic reference frames.}
\label{RMI}
\end{figure} 
\noindent not only gravitational forces, but
also non-gravitational ones. In other words, it is those 4-dimensional
equations (i.e. $\frac{d^{2}X^{\alpha }}{d\tau ^{2}}$) which are directly
numerically integrated. This new approach is called (SC)RMI
((Semi-Classical) Relativistic Motion Integrator). RMI was first suggested by X. Moisson 
and S. Loyer during their PH.D. Thesis at the Observatoire de Paris \cite{these Moisson}. 
Figure \ref{RMI} summarizes the (SC)RMI approach.\newline

For the appropriate metric at the required order (power series in $1/c^2$), the general relativistic
equations (\ref{generalized_equation_of_motion}) contain all the
gravitational effects at the corresponding order. They are coded in the
Christoffel symbols that are functions of the derivative of the metric. The
metric $G^{\alpha \beta }$ itself encodes the geometry of space-time that
is shaped (deformed) by the presence of solar system bodies. 
The above equations (\ref{generalized_equation_of_motion}) computed for 
the Geocentric Coordinate Reference System (GCRS) metric \cite{IAU2000 resolutions}, \cite{explanatory on IAU2000 resolutions} will take into account gravitational multipole moment
contributions from the central planetary gravitational potential,
perturbations due to solar system bodies, the Schwarzschild, geodesic and
Lense-Thirring precessions.\newline
\indent Non-gravitational forces can be treated as
perturbations, in the sense that they do not modify the local structure of
space-time (the metric). Moreover, $K_{\beta }$ being small, one can safely
replace $G^{\alpha \beta }$ by its Minkowskian counterpart, $\eta^{\alpha \beta }$, 
in the second term of the right-hand-side of equation (\ref{generalized_equation_of_motion}), 
hence the terminology ``Semi-Classical'' in SCRMI.
This leads to 
\begin{equation}
\frac{dU^{\alpha }}{d\tau }=-\Gamma _{\beta \gamma }^{\alpha }U^{\beta
}U^{\gamma }+K_{\beta }\left( \eta^{\alpha \beta }-\frac{U^{\alpha }}{c}%
\frac{U^{\beta }}{c}\right)   \label{bis_generalized_equation_of_motion}
\end{equation}
\smallskip 

In the classical limit, equations (\ref{generalized_equation_of_motion}, \ref{bis_generalized_equation_of_motion})
reduce to the Newtonian gravitational 3-acceleration equation (\ref{newton_equation_of_motion})
(without relativistic corrections) where $W$ is a generalized gravitational
potential given in the GCRS metric.

\subsection {A symplectic integrator}

Expression (\ref{generalized_equation_of_motion}) consists in fact of four
equations, to compare with the three equations (\ref{newton_equation_of_motion})
to be integrated in Newtonian mechanics. However, a first integral exists, 
i.e. equation (\ref{norm_conservation_equation}), since the norm (with respect to
the metric $G^{\alpha \beta }$) of the quadri-proper-velocity, $U^{\alpha }$, 
is conserved along the trajectory. This point stresses the importance of a
symplectic integrator naturally preserving this quantity.

\subsection{(SC)RMI validation}

(SC)RMI, based on the relativistic equations (\ref{bis_generalized_equation_of_motion}), 
is validated by comparison with template orbits from GINS in which a particular 
relativistic effect has been switched on. 
For example, using, in (SC)RMI, the Schwarzschild metric (based on a
single central spherical static body) allows to validate the Schwarzschild
precession. If a multipolar geopotential model is added to the Schwarzschild
mass monopole, one can further validate the harmonic contributions
of the potential. Using the Kerr metric (based on a single central spherical
rotating gravitational body) allows to validate an additional relativistic
effect: the Lense-Thirring precession. Finally, using the GCRS metric taking into account the
Solar and planetary gravitational fields will include additional effects
like the geodesic precession and the perturbations from the corresponding planets.%
\newline
Once completed, (SC)RMI will go beyond GINS, including the IAU2000/IERS 2003 new
standards regarding metric, space-time transformations and Earth rotation
models. Moreover, separate modules allow easy updates for metric, Earth
potential... IAU recommendations, keeping the integrator body unchanged. It
is also important to stress that the (SC)RMI approach natively contains not only
all the relativistic effects at the corresponding order of the metric, but
also all the couplings between these effects at the corresponding order.

\section {The principle of accelerometers}

Last we show how to update the classical equation for accelerometers; in
other words, how to measure $K_{\beta }$, or how to introduce a
non-gravitational force model in the relativistic framework.

\noindent Let the satellite center of mass (CM) be located at $X^{\mu }$; while a test-mass is
at $X^{\mu }+\delta X^{\mu }$, in a cavity inside the satellite, hence
shielded from non-gravitational forces. The test-mass motion is described by
geodesic equations ((\ref{generalized_equation_of_motion}) with $K_{\beta }$%
=0) while that of the satellite is described by (\ref
{generalized_equation_of_motion}). Evaluating the difference between those
two equations at first order in $\delta X^{\mu }$ gives a general
relativistic equation for accelerometer measurements: 
\begin{equation}
\frac{d^{2}\delta X^{\alpha }}{d\tau ^{2}}=\left. 
\begin{array}[t]{l}
+K_{\beta }^{(CM)}\left( G^{\alpha \beta }-\frac{dX^{\alpha }}{d\tau }\frac{dX^{\beta }}{d\tau }\right)\smallskip \\ 
-\frac{\partial \Gamma _{\beta \gamma }^{\alpha }}{\partial X^{\mu }}\delta X^{\mu }%
\frac{dX^{\beta }}{d\tau }\frac{dX^{\gamma }}{d\tau }\smallskip \\ 
-2\Gamma _{\beta \gamma}^{\alpha }\frac{dX^{\beta }}{d\tau }\frac{d\delta X^{\gamma }}{d\tau }
\end{array}
\right.  
\label{generalized_geodesic_deviation}
\end{equation}
Equation (\ref{generalized_geodesic_deviation}) reduces to geodesic
deviation if $K_{\beta }^{(CM)}$=0.\newline
In the classical limit, one recovers the classical accelerometer equation,
\begin{equation}
\frac{d^{2}\delta X^{i}}{dT^{2}}=
-K_{i}^{(CM)}
-{\sum}_{j} {\frac{\partial ^{2}W}{\partial X^{i}\partial X^{j}}\delta X^{j}}
\end{equation}
where $\frac{d^{2}\delta X^{i}}{dT^{2}}$ is the 3-dimensional
relative acceleration of the satellite with respect to the test mass; the
third term of equation (\ref{generalized_geodesic_deviation}) vanishes
completely.
\vspace*{0.25cm}

\section {Conclusions}

We have outlined here a new paradigm for orbitography software that is, in our opinion, a key for the future. The concepts differs radically from those in use today. To implement this idea requires rewriting the core parts of existing programs; they cannot be simply ``upgraded". This is obviously the main difficulty to overcome.
The (SC)RMI prototype software can be validated using a classical orbitography software thanks to the progressive method described in Section 3.4. Some of step of this method have already been implemented. In reference ``Non-gravitational forces and the relativistic equations of motion" \cite{Pireaux 2006 non-grav forces} in preparation, the authors provide further details on symplectic integrators (mentioned in Section 3.3 of the present article) and on the non-gravitational force term (to be measured from accelerometers as described in Section 4, or provided by classical models).

\vspace*{0.5cm}
\noindent\large {\bf Acknowledgements}
\vspace*{0.5cm}

\small {S. Pireaux acknowledges financial support provided through the
European Community's Improving Human Potential Program under contract
RTN2-2002-00217, MAGE. JP. Barriot is funded by CNES/INSU. P. Rosenblatt is
funded under a Belgian ESA/\newline 
PRODEX contract.}



\begin{thebibliography}{9}
\bibitem{descriptif GINS}  {\tiny \ }GRGS, Descriptif mod\`{e}le de forces:
logiciel GINS. Note technique du Groupe de Recherche en G\'{e}odesie
Spatiale (GRGS), 2001.

\bibitem{IAU2000 resolutions}  IAU 2000 resolutions, IAU Information
Bulletin 88 (2001).\\
Erratum on resolution B1.3. Information Bulletin 89 (2001).

\bibitem{explanatory on IAU2000 resolutions}  M. Soffel, S. A. Klioner, G. Petit, P. Wolf, S. M. Kopeikin, P. Bertagon, V. A. Brumberg, N. Capitaine, T. Damour, T. Fukushima, B. Guinot, T. Huang, T. Lindegren, C. Ma, K. Nordtvedt, J. Ries, 
P. K. Seidelmann, D. Vokrouhlicky, C. Will and C. Xu, 
The IAU 2000 resolutions for astrometry, celestial mechanics and metrology in the
relativistic framework: explanatory supplement, 
Astronomical Journal 126, 6 (2003), p. 2687-2706. astro-ph/0303376v1.

\bibitem{these Moisson}   X. Moisson, Int\'{e}gration du mouvement des plan\`{e}tes dans 
le cadre de la relativit\'{e} g\'{e}n\'{e}rale (th\`{e}se). Observatoire de Paris, 2000.

\bibitem{Lichnerowicz 1995 RG and em}  A. Lichnerowicz, Th\'{e}ories
relativistes de la gravitation et de l'\'{e}lectromagn\'{e}tisme, Masson et Cie Editeurs 1955.

\bibitem{Pireaux 2006 non-grav forces} S. Pireaux, J-P. Barriot, Non-gravitational forces and the relativistic equations of motion, in preparation for Celestial Mechanics (2006).
\end{thebibliography}
\end{document}